# FEAorta: A Fully Automated Framework for Finite Element Analysis of the Aorta From 3D CT Images

Jiasong Chen[a], Linchen Qian[a], Ruonan Gong[a], Christina Sun[b], Tongran Qin[b], Thuy Pham[b], Caitlin Martin[b], Mohammad Zafar[c], John Elefteriades[c], Wei Sun[b], Liang Liang*[a]

[a] Department of Computer Science, University of Miami, Coral Gable, FL, USA

[b] Sutra Medical Inc, Lake Forest, CA, USA

[c] Aortic Institute at Yale-New Haven Hospital, Yale University School of Medicine, New Haven, CT, USA

## 1. Introduction

Aortic aneurysm disease ranks consistently in the top 20 causes of death in the U.S. population [1]. Thoracic aortic aneurysm (TAA) is manifested as an abnormal bulging of thoracic aortic wall and it is a leading cause of death in adults [2]. TAA prevalence is relatively high, estimated to be around 1% of the general population [3,4]. TAA is a silent and sudden killer. The progression of TAA is a silent process, yet rupture can often occur suddenly, without any premonitory signs or symptoms: for many TAA patients, the first symptom is often death [5]. Surgical repair can prevent these aortic events. The current criterion for surgical intervention is based on aneurysm diameter. It is almost unanimously viewed that surgical intervention is sensible when TAA diameter > 5.5 cm indicating high risk of rupture [6,7]. However, the diameter-based criterion cannot assess risk of smaller TAAs which may or may not rupture [2,8,9], and this criterion is close to random guess for TAAs with diameters ≤ 5cm [10,11]. It is estimated that there are millions of patients in the U.S. with smaller TAAs ≤ 5 cm [8], and the yearly rupture (full, partial, death) rate is ~7% [8], who are ignored by the current diameter-based criterion. If the high-risk TAA patients (i.e., the 7%) can be identified, their lives can be improved/saved by surgeries.

From the perspective of biomechanics, rupture occurs when the stress acting on the aortic wall exceeds the wall strength [12,13]. Wall stress distribution can be obtained by computational biomechanical analyses, especially structural Finite Element Analysis (FEA). For risk assessment, probabilistic rupture risk of TAA can be calculated by comparing stress with material strength using a material failure model [10,14]. Although these engineering tools are currently available for TAA rupture risk assessment on patient-specific level, clinical adoption has been limited due to two major barriers: (1) labor-intensive 3D reconstruction – current patient-specific anatomical modeling still relies on manual segmentation, making it time-consuming and difficult to scale to a large patient population, and (2) computational burden – traditional FEA simulations are resource-intensive and incompatible with time-sensitive clinical workflows.

The second barrier was successfully overcome by our team through the development of the PyTorch-FEA library [15] and the FEA–DNN integration framework [16]. By incorporating the FEA functionalities within PyTorch-FEA and applying the principle of static determinacy, we reduced the FEA-based stress computation time to approximately three minutes per case. Moreover, by integrating DNN and FEA through the PyTorch-FEA library, our approach further decreases the computation time to only a few seconds per case.

This work focuses on overcoming the first barrier through the development of an end-to-end deep neural network (DNN) capable of generating patient-specific finite element meshes of the aorta directly from 3D CT images.

## 2. Related Work

Leveraging the principles of real-world physical laws, computational simulations of the aorta serve as powerful tools for investigating hemodynamic behavior and structural mechanics. These simulations play a pivotal role in advancing cardiovascular research and supporting clinical decision-making, enabling noninvasive assessment of disease progression, surgical planning, and device evaluation. Finite element (FE) simulations provide a detailed and quantitative approach to evaluate the distribution of stress and strain throughout the aortic wall. By integrating patient-specific anatomical and physiological parameters—such as variations in wall thickness, blood pressure, and vessel geometry—FE simulations enable the patient-specific aortic biomechanical behavior investigation of how the aorta responds to varying mechanical loads under realistic physiological conditions. FE simulations are instrumental in aneurysm risk assessment as well as modeling aneurysm growth and disease progression over time. In particular, wall stress distributions derived from patient-specific geometries can identify sites of potential rupture beyond what conventional diameter-based evaluation alone can indicate [17–20]. FE simulations facilitate fluid-structure interaction (FSI) studies by supplying aortic wall mechanical properties, thereby enabling a comprehensive analysis of the coupled dynamics between pulsatile blood flow, pressure variations and aortic wall displacement [21,22]. In addition, FE simulations support the estimation of individualized material properties of the human aorta, enhancing patient-specific modeling by determining individual nonlinear and anisotropic aortic wall characteristics [23,24]. Furthermore, FE simulations play an essential role in cardiovascular intervention planning and surgical optimization by reconstructing patient-specific vascular geometries and simulating biomechanical interactions between tissues and implanted devices, enabling stress–strain analysis to predict postoperative patient outcomes and assist in developing personalized treatment strategies [25,26].

The widespread adoption of FE simulation in cardiovascular research depends on two critical factors: usability and accuracy, both of which are closely linked to the quality of the mesh representing the target anatomy. Conventional workflows typically involve multiple labor-intensive steps, including image segmentation, mesh generation, and simulation [27–30]. Segmentation is used to extract volumetric masks from medical images, representing the anatomical shape of the target structure. However, segmentation outputs often contain artifacts and irregularities that make them unsuitable for direct use in biomechanical simulations. Therefore, a meshing step is required to convert the segmentation masks into high-quality, simulation-ready meshes. FE simulation then solves the governing partial differential equations (PDEs) on these discretized mesh elements, approximating the continuum mechanics of the aorta and enabling replication of its real-world biomechanical behavior under physiological loading conditions. Streamlining these processes while maintaining high simulation fidelity is essential for enabling routine clinical and research applications.

Manual segmentation and meshing require specialized medical knowledge and meshing expertise, making the process labor-intensive and time-consuming. Simplifying the mesh generation workflow would greatly enhance the usability of FE simulation tools. In addition to rapid mesh generation, producing high-quality meshes is essential for accurate and efficient simulations. The definition of "high-quality" varies depending on the application. From an accuracy perspective, the mesh must closely capture the detailed anatomical geometry of the target structure. From an efficiency perspective, the mesh should contain an appropriate number of evenly distributed nodes to avoid unnecessary computational cost. A high-quality mesh strikes a balance between these requirements to optimize simulation outcomes. Recent machine learning approaches have demonstrated that direct image-to-mesh methods can generate high-quality meshes without requiring intermediate segmentation steps, significantly reducing processing time and dependency on expert input.

Machine learning has emerged as powerful tool in medical image analysis. For example, Francesca *et al*. developed a multi-view U-Net architecture to localize and segment the abdominal aortic aneurysms (AAA) thrombus from the contrast-enhanced computed tomography angiography (CTA) images [31]. The Segment Anything Model (SAM) was originally designed for natural image segmentation, demonstrating high adaptability with minimal or no additional training. Recent studies have extended SAM to volumetric medical image segmentation tasks, particularly for aortic applications. Wang *et al.* proposed SAM-Med3D, a model trained on a large-scale volumetric medical dataset for general-purpose 3D medical image segmentation [32]. Iltaf *et al.* introduced VesselSAM, which integrates the AtrousLoRA module, a combination of Atrous Attention and Low-Rank Adaptation (LoRA), to enhance SAM's performance in aortic vessel segmentation [33]. Zohranyan *et al.* further expanded SAM's capability through a novel positive point selection strategy combined with user-defined bounding boxes, enabling more refined vascular segmentation [34]. Cai *et al.* developed LABEL-SAM, a semi-automatic interactive segmentation algorithm for annotating aortic dissections in 3D CTA images, requiring only minimal user input through points and bounding boxes on selected slices [35].

In parallel, research on mesh generation has increasingly focused on direct image-to-mesh translation for simulation-ready geometries. For example, Voxel2Mesh takes a volumetric medical image as input and deforms an initial spherical mesh into a target 3D surface mesh [36]. MeshDeformNet deforms a predefined mesh template to directly generate a whole-heart surface mesh from volumetric medical images [37]. However, such approaches may produce irregular mesh structures, including elements with high skewness or self-intersections, which make them unsuitable for FE simulations. To address these issues, incorporating diffeomorphic constraints into template deformation-based methods can ensure smooth and topologically consistent mesh generation. By regularizing the deformation field on a template mesh, these methods prevent irregular element shapes and preserve mesh integrity. For instance, CorticalFlow predicts a dense 3D flow field from volumetric images to smoothly deform a mesh template while maintaining anatomical consistency [38]. Similarly, Pak *et al.* proposed learning a diffeomorphic B-spline deformation field to deform mesh templates without self-intersections, ensuring geometric smoothness and preserving structural fidelity [39].

Using a predefined structured mesh template also ensures mesh correspondence, which is critical for downstream tasks such as shape analysis and landmark estimation. For anatomical analyses, the aorta is typically subdivided into five main segments: the aortic root, ascending thoracic aorta, aortic arch, descending thoracic aorta, and abdominal aorta. Beyond the standard anatomical description, the concept of aortic landing zones divides the thoracic and abdominal aorta into 11 regions, providing a technical framework for planning, guiding, and reporting aortic interventions, particularly endovascular stent-grafting [40–42]. Each anatomical system offers distinct advantages for characterizing aortic geometry. In previous studies, simulations were either performed on specific aortic segments or on the entire aorta, followed by manual post-processing required to analyze results for individual anatomical regions [43–45]. Simulations of the entire aorta provide more comprehensive and precise biomechanical information than segment-specific analyses, but manual post-processing is time-consuming and inconsistent. This limitation can be addressed using a structured mesh template, as the built-in mesh correspondence enables automated evaluation of stress, strain, and other simulation outputs across defined anatomical sections [30].

To address the challenges of finite element analysis of the aorta, we propose a fully automated framework for patient-specific FEA directly from 3D CT images. At the core of this framework is a deep learning-based image-to-mesh template deformation method that generates high-quality meshes while maintaining consistent mesh correspondence. This approach overcomes key barriers in 3D aortic reconstruction from CT images, including rapid mesh generation, mesh quality assurance, and region-specific mesh fidelity. By integrating the PyTorch-FEA library to handle computational demands of simulation, our framework establishes a fully automated and comprehensive pipeline for patient-specific biomechanical analysis of the aorta.

3. **Methodology**

Traditional approaches typically divide the simulation workflow into three separate stages (segmentation, meshing, simulation), which complicate FE simulations and limit their usability and efficiency. An alternative approach—directly predicting stress fields from images—requires learning a surrogate model that maps geometrical and boundary condition information encoded in the images to continuous simulation outputs while maintaining physical consistency. Such an approach demands complex network architectures to capture the high-dimensional relationship between input and output [46]. To balance complexity, interpretability, and performance, we adopt a two-stage strategy: meshing and simulation. In the meshing stage, the model takes a 3D CT image as input and deforms a predefined quadrilateral template to generate a patient-specific, high-quality quadrilateral mesh suitable for simulation. In the simulation stage, stress computations are performed directly on the quadrilateral mesh produced in the first stage. Figure 1 illustrates the overview of our proposed framework.

Our goal is to directly extract anatomical information from the target image and use it to deform a quadrilateral mesh template into a patient-specific quadrilateral mesh corresponding to the target anatomy. Specifically, 3D CT images are input into the SVF network (SAM-SVF), which predicts stationary velocity field. This stationary velocity field is then converted into a displacement field by the DMD module, which is subsequently applied to the template mesh to generate the patient-specific mesh. Finally, biomechanical stress analysis is performed on the predicted mesh using the PyTorch-FEA library.

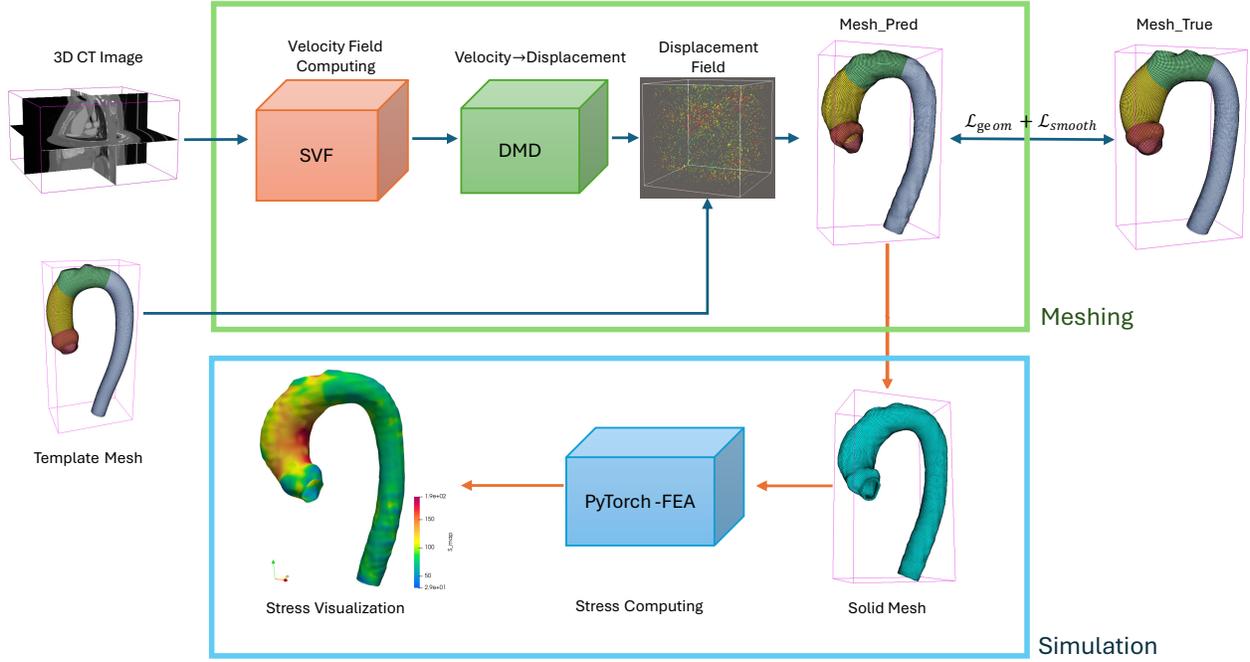

Figure 1. Overview of FEAorta Framework

### 3.1. Problem Formulation

Let the quadrilateral mesh be $\mathcal{M} = (V, F)$ with vertices V and face elements $F$. Here, $V = \{v_i \epsilon \mathbb{R}^{n \times 3}\}_{i=1}^{|V|}$ with $|V|$ is the number of vertices. $F = \left\{ f_k = (v_{k_1}, v_{k_2}, v_{k_3}, v_{k_4}) \middle| v_{k_j} \epsilon V, \; j = 1, \dots, 4 \right\}_{k=1}^{|F|}$ with $|F|$ is the number of quadratic face elements. Each face element $f_k$ is an ordered 4-tuple of vertices from V. The objective of mesh deformation is to estimate the optimal displacement field ($\delta$) that transforms the mesh template $\mathcal{M}_0 = (V_0, F_0)$, to the target mesh $\mathcal{M}_T = (V_T, F_T)$, which involves directly or implicitly estimate the displacement vector ($u_i$) at every vertex $i$ of mesh template. $u_i$ also has another notation as $\delta(v_i)$.

$$\delta = \{u_i = v_i^T - v_i^0 \mid v_i^0 \in V_0, v_i^T \in V_T\}_{i=1}^{|V|}$$

Then, the optimal target mesh could be derived by minimizing the deformation loss ($\mathcal{L}$).

$$\mathcal{M}_T^* = \mathcal{M}_0 + \arg\min_\delta \mathcal{L}(\mathcal{M}_T, \mathcal{M}_0, \delta)$$

The displacement field ($\delta$) can be estimated implicitly by first computing an embedded stationary velocity field ($\tau$) and subsequently integrating it into the displacement field [47]. Under this implicitly displacement estimation with embedded stationary velocity field, the displacement field can smoothly deform the mesh template. The details of the diffeomorphic module ($\Gamma$) that converts stationary velocity field to displacement field will be further explained.

$$\delta(v_i) = \phi(v_i) - v_i, \text{ with } \phi = \exp(\tau)$$

Given a volumetric image $I \epsilon \mathbb{R}^{H \times W \times D}$ with height $H$, width $W$ and depth $D$, a neuron network $\psi_\theta$ parameterized by $\theta$ is used to predict the optimal stationary velocity field. Then, the optimal corresponding target mesh $\mathcal{M}_I^*$ within target volumetric image $I$ can then be obtained by optimizing $\theta$ to minimize the deformation loss over the paired target images $I$ and meshes $\mathcal{M}_I$ in training dataset.

$$\theta^* = \arg\min_\theta \mathcal{L}(\mathcal{M}_I, \mathcal{M}_0, \psi_\theta(I))$$
$$\mathcal{M}_I^* = \mathcal{M}_0 + \Gamma(\psi_\theta(I))$$

### 3.2. Template Generation

To construct a robust and representative quadrilateral mesh template for deformation, all quadrilateral meshes in the training dataset are aligned based on the established mesh correspondences. For each corresponding node across these meshes, we computed the average position to generate a single, representative quadrilateral mesh. This average mesh effectively captures the typical aorta geometry while preserving essential anatomical features, such as curvature, branching patterns, and wall contours. By serving as a standardized reference, the template provides a smooth and anatomically meaningful initial geometry that can be efficiently deformed to match patient-specific anatomy. This approach not only ensures consistency across training samples but also facilitates reliable learning for deformation models, improving the robustness and accuracy of mesh predictions in subsequent stages.

### 3.3. SAM_SVF

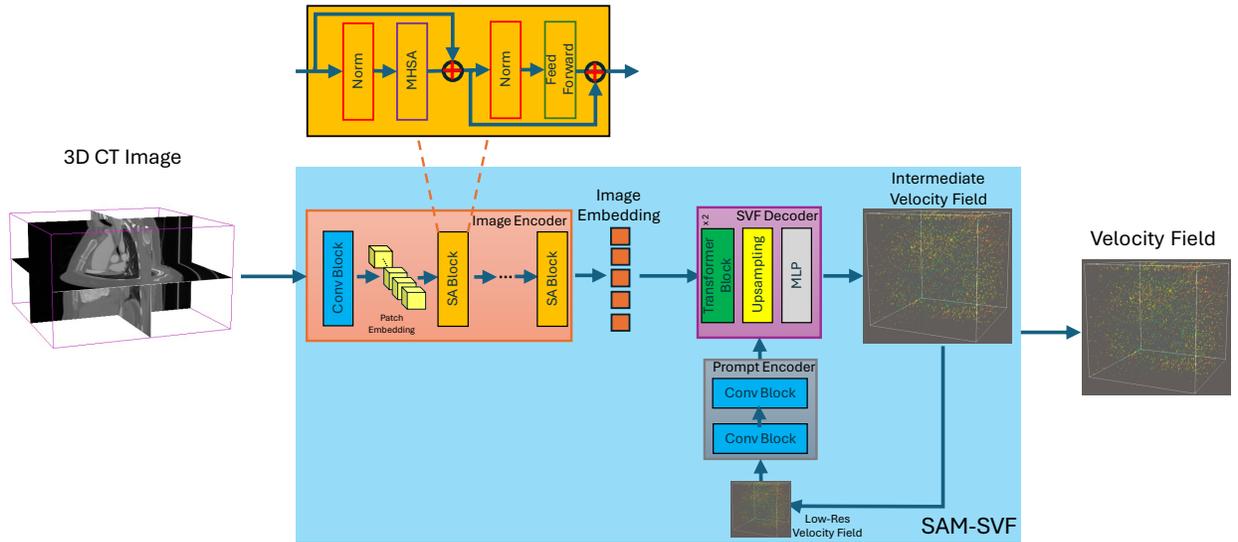

Figure 2. Overview of SAM-SVF

In general, the SVF network module in our framework can employ any architecture, provided that the input is a 3D CT image and the output is a stationary velocity field (SVF). In this work, we introduce SAM-SVF, a network specifically designed to predict SVFs from medical images. SAM-SVF comprises three main learnable components: the Image Encoder, the Prompt Encoder, and the Stationary Velocity Field (SVF) Decoder. Figure 2 shows the architecture of SAM-SVF.

The Image Encoder employs a hybrid CNN-Transformer architecture to extract rich visual features from the input medical images, producing a high-dimensional volumetric image embedding. First, the 3D CT image is processed through a 3D CNN block to extract feature maps. This CNN block is implemented as a residual block consisting of three convolutional filters with shortcut connections. The resulting feature maps are then fed into a Vision Transformer (ViT) [48], which performs patch embedding and applies self-attention layers to capture global contextual information. The final output of the Image Encoder is a volumetric image embedding that encodes both local and global structural features of the input anatomy.

The Prompt Encoder processes a lower-resolution stationary velocity field through two convolutional blocks, converting it into compact feature vectors that serve as deformation guidance for the network. These prompt embeddings provide prior information about the expected deformation, enabling SAM-SVF to efficiently capture complex correspondences between the input 3D image and the mesh template. By integrating both local and global deformation cues, the Prompt Encoder helps the network generate smoother and more anatomically consistent stationary velocity fields. This, in turn, enhances the accuracy and robustness of the predicted SVF, ensuring that the resulting patient-specific mesh accurately reflects the target anatomy.

The SVF Decoder is a transformer-based module that fuses the image embedding and prompt features to generate the SVF. The decoder begins by concatenating the feature maps from the Image Encoder and Prompt Encoder and processing them through two transformer layers. These layers update the image embedding using self-attention on the prompt embedding and bidirectional cross-attention between the image embeddings and prompt embeddings, allowing the lower-resolution SVF context to effectively guide the deformation task. Next, the updated image embedding is upsampled by a factor of four using two transposed convolution layers. Simultaneously, the updated prompt embedding is passed through a three-layer MLP. The upscaled image embedding is then combined with the MLP output via a spatially point-wise product, followed by tri-linear interpolation to match the resolution of the original input image, producing the high-resolution SVF. To enhance deformation alignment, the three dimensions of the output SVF are treated separately as individual "classes", enabling precise mapping between the template mesh and the target anatomy. The intermediate output SVF is also interpolated to a lower resolution and reused as input to the Prompt Encoder in the next iteration. In each loop, the decoder leverages the lower-resolution SVF from the previous iteration and fuses it with the image embedding to generate a refined SVF, progressively improving deformation accuracy and capturing fine anatomical details. After several iterations, the SVF Decoder produces the final, high-resolution SVF that accurately encodes the detailed mesh deformation.

### 3.4. Loss Function

The loss function combines a mesh geometric loss and a smoothness loss. The mesh geometric loss is defined as the mean squared error (MSE) between the ground truth mesh vertices and the predicted mesh vertices. Since different anatomical regions of the aorta contain varying numbers of mesh points, training the network with an unweighted geometric loss could cause it to neglect finer details in regions with fewer points. To address this, we construct a weighted geometric loss that guides the network to learn surface details across all regions.

Specifically, the aortic mesh domain ($\Omega$) is divided into four non-overlapping anatomical regions: the aortic root ($\Omega_{root}$), ascending aorta ($\Omega_{ascending}$), aortic arch ($\Omega_{arch}$), and descending aorta ($\Omega_{descending}$). The weighted geometric loss computes a region-specific mean squared error and aggregates them using region-specific weights, ensuring balanced contribution from each anatomical region during training. This allows the network to capture detailed surface geometry uniformly across the entire aorta.

$$\Omega = \Omega_{root} \cup \Omega_{ascending} \cup \Omega_{arch} \cup \Omega_{descending}, \Omega_i \cap \Omega_j = \emptyset \ (i \neq j)$$

$$\mathcal{L}_r = \frac{1}{|\Omega_r|} \sum_{v \in \Omega_r} \|\hat{v} - v\|^2$$

$$\mathcal{L}_{weighted_{geo}} = \sum_r \omega_r \mathcal{L}_r, \sum_r \omega_r = 1, \omega_r \geq 0$$

The smoothness loss is defined on mesh edges and penalizes large angles between adjacent edges, encouraging smoother surface mesh. A hyperparameter ($\alpha$) controls the strength of the smoothing effect. The total loss used during training is a combination of the weighted geometric loss and the smoothness loss, with an additional hyperparameter to balance their relative contributions. This formulation ensures that the network not only accurately captures the mesh geometry but also produces smooth, simulation-ready meshes suitable for simulation.

$$\mathcal{L}_{total} = \mathcal{L}_{weighted_{geo}} + \alpha \mathcal{L}_{smooth}$$

## 4. Experiments

### 4.1. Dataset

Our datasets were obtained from the INSPECT dataset [49], a multimodal medical dataset comprising data from 19,402 patients for pulmonary embolism diagnosis and prognosis. For this study, we randomly selected 100 non-aneurysmal patients and 55 aneurysmal patients. Three experts independently performed manual annotation of segmentations and anatomical landmarks using 3D Slicer [50]. The annotated landmarks included three hinge points and three commissure points.

The ground truth quadrilateral meshes were generated following the established quadrilateral mesh generation pipeline [30]. Each patient-specific quadrilateral mesh contains 24,960 nodes and 24,882 face elements. For model training, 75 non-aneurysmal and 30 aneurysmal patients were randomly selected, while the remaining 25 non-aneurysmal and 25 aneurysmal patients were used as the testing dataset.

### 4.2. Implementation Details

We preprocess the 3D CT images by first normalizing pixel intensities to the range [0, 1]. All images are then resampled to an isotropic spatial resolution of 1 mm³. Next, the images are cropped around the center of the labeled aortic geometries using a bounding box that is 1.1× the size of the smallest aorta bounding box. The cropped images are subsequently isotropically scaled along the diagonal axes until at least one dimension (height, width, or depth) reaches 256 voxels. After padding the shorter sides, all images are resized to a uniform volume of [256, 256, 256]. During training, data augmentation techniques—including translation, rotation, and elastic deformation—are applied to enhance model generalization.

### 4.3. Evaluation Metrics

For the meshing experiments, we evaluated the meshing performance using the chamfer distance, which measures the average point-to-point distance between the predicted mesh and the ground truth mesh. For the simulation experiments, we assessed performance using mean error metric. The mean error is computed by averaging the percentage differences in mean stress across mesh elements in different anatomical regions of aorta. These metrics provide a quantitative assessment of both mesh reconstruction accuracy and the fidelity of subsequent finite element simulations.

## 5. Results and Discussion

This section represents experimental results including mesh quality evaluation, geometric accuracy assessment, simulation performance and clinical implication analysis. Since the mesh quality significantly impacts the stability and accuracy of finite element simulation, we first report the results of mesh quality. Next, we assess geometric accuracy to evaluate the network's performance. Subsequently, we present the simulation results, followed by an analysis of the clinical relevance results and findings. We also discuss the implications of these findings for clinical applicability and future research directions. The results are analyzed in detail to highlight methodological advantages and potential limitations, providing a comprehensive evaluation of the proposed framework.

### 5.1. Mesh Quality Evaluation

Mesh element quality plays a critical role in determining the stability and accuracy of finite element simulations. We evaluated the quality of the generated mesh elements primarily using the VTK [51]. Table 1 summarizes the mesh quality across several metrics, including Equiangle Skew, Scaled Jacobian, Aspect Ratio, Minimum/Maximum Angle and Self-Intersection. The error for each metric was calculated as the average element-wise difference between the predicted mesh and the ground truth mesh. These six metrics provide complementary measures of how closely the predicted mesh elements approximate the ideal mesh element, assessing geometric regularity, distortion, and element shape from multiple perspectives.

Equiangle Skew ranges from 0 to 1, with lower values indicating well-shaped mesh elements and measuring how much the element angles deviate from the ideal equiangular shape. The predicted mesh elements have a mean equiangular skew of 0.0989, compared with 0.0592 for the ground truth mesh. The Aspect Ratio of a planar quadrilateral ranges from 1 to infinity, with smaller values being better. The mean aspect ratio of the predicted mesh is 1.1642, compared with 1.1381 for the ground truth, where the ideal aspect ratio is 1.

The Minimum and Maximum Angle metrics show that the quadrilateral angles of the predicted mesh range from 81.49° to 98.55°, close to the ideal 90° for quadrilateral elements. The Scaled Jacobian, which ranges from -1 to 1, measures the transformation from an ideal element to the target element, with higher values indicating near-perfect, orthogonal elements. The predicted mesh has a mean scaled Jacobian of 0.981, compared with 0.993 for the ground truth.

Overall, these results demonstrate that the quadrilateral meshes generated by our proposed method exhibit excellent quality across multiple metrics. High-quality meshes are essential for ensuring accurate and stable finite element simulations, significantly improving the reliability of simulation results.

Table 1. Mesh Quality Results

|  | EQUIANGLE SKEW | ASPECT RATIO | SCALED JACOBIAN | MIN ANGLE | MAX ANGLE | SELF-INTERSECTION |
|---|---|---|---|---|---|---|
| Ground Truth | 0.0592 ± 0.0066 | 1.1381 ± 0.0272 | 0.993 ± 0.0019 | 84.9861 ± 0.5749 | 95.0428 ± 0.5848 | 0 |
| Prediction | 0.0989 ± 0.0124 | 1.1642 ± 0.0237 | 0.981 ± 0.0049 | 81.4917 ± 1.0745 | 98.5458 ± 1.0760 | 0 |
| Error | 0.0397 ± 0.0150 | 0.0299 ± 0.0225 | 0.012 ± 0.0054 | 3.4944 ± 1.2890 | 3.503 ± 1.2950 | 0 |

### 5.2. Mesh Geometric Evaluation

We adopted chamfer distance to evaluate the performance of generating quadrilateral meshes directly from 3D CT images. Table 2 reports the chamfer distance of meshes across different anatomical regions of the aorta, quantifying the geometric differences between the predicted meshes and the ground truth. The mean chamfer distance for the full aorta mesh across all patients was 1.4889 mm. It is important to note that our predicted meshes are extracted directly from volumetric CT images, with a default voxel spacing of 1 mm × 1 mm × 1 mm. Compared with traditional methods that reconstruct meshes via marching cubes from segmentation, our predicted mesh nodes may locate at any position within a 1 mm³ voxel. The theoretical maximum distance between two nodes within the same voxel corresponds to the voxel diagonal, $\sqrt{3}mm$. This demonstrates that our predicted meshes demonstrate high fidelity in representing complex anatomical geometry.

In addition, the Chamfer distance error for aneurysmal patients was slightly higher than that for non-aneurysmal patients, likely due to larger and more complex deformations from the template mesh. When comparing Chamfer distances across different regions of the aorta, the aortic root exhibited the largest mean error, reflecting its intricate cloverleaf-like geometry, whereas the other three sections are relatively closer to simple cylindrical shapes.

Table 2 Mesh Chamfer Distance Results

|  | Aortic Root | Ascending Aorta | Aortic Arch | Descending Aorta | Full Aorta |
|---|---|---|---|---|---|
| Non-Aneurysm | 1.7074 ±0.6614 | 1.3936 ±0.4402 | 1.3998 ±0.5415 | 1.4766 ±0.5238 | 1.4011 ±0.3916 |
| Aneurysm | 2.0116 ±0.7681 | 1.7363 ±0.6157 | 1.8951 ±0.5186 | 1.6299 ±0.5613 | 1.5767 ±0.4097 |
| Total | 1.8595 ±0.7327 | 1.5649 ±0.5620 | 1.6475 ±0.5852 | 1.5532 ±0.5483 | 1.4889 ±0.4102 |

### 5.3. Simulation Evaluation

Due to geometric differences between non-aneurysmal and aneurysmal aortas, differences in stress results are clearly observed. Table 3 compares the mean stress errors across different anatomical regions for both patient types. These errors reflect the simulation performance of the predicted meshes. Overall, the mean stress errors in aneurysmal patients are slightly higher than those in non-aneurysmal patients across all regions, except for the aortic arch. This can be attributed to the generally larger geometry of aneurysmal aortas, which leads to higher stress under the same loading conditions, thereby slightly increasing the corresponding simulation errors.

Furthermore, we observed that the predicted meshes in the descending aorta often exhibit a staircase effect, which affects the simulation results and contributes to higher stress errors in this region. In contrast, the aortic root, ascending aorta, and aortic arch exhibit lower stress errors, demonstrating the effectiveness and accuracy of our proposed network in capturing complex anatomical geometries.

Table 3 Simulation Results

|  | **Aortic Root** | **Ascending Aorta** | **Aortic Arch** | **Descending Aorta** |
|---|---|---|---|---|
| **Non-Aneurysm** | 0.072 ±0.061 | 0.050 ±0.043 | 0.072 ±0.053 | 0.113 ±0.092 |
| **Aneurysm** | 0.088 ±0.065 | 0.055 ±0.042 | 0.061 ±0.053 | 0.157 ±0.148 |
| **Total** | 0.080 ±0.064 | 0.053 ±0.043 | 0.066 ±0.053 | 0.135 ±0.125 |

## 5.4. Clinical Implication on Population Level

In this section, we present clinically relevant results. Leveraging the high quality of the quadrilateral mesh with mesh correspondence, our approach enables the measurement of maximum cross-sectional diameters and stress distribution along different regions of the aorta. The maximum cross-sectional diameter is a key clinical indicator used to assess a patient's aortic status, aiding in the distinction between non-aneurysmal and aneurysmal aortas and guiding clinical decision-making, particularly for aneurysmal patients.

Table 4 summarizes the differences in maximum cross-sectional diameters between the ground truth and predicted meshes. Across all anatomical regions, aneurysmal patients exhibited larger maximum diameters than non-aneurysmal patients, with the ascending aorta showing the most pronounced difference. Comparison across different aortic regions further highlights that the ascending aorta of aneurysmal patients has the largest cross-sectional diameter, a finding accurately captured and validated by our method. For the ascending aorta, the average error across all patients was 1.706 mm, with aneurysmal patients showing slightly higher errors than non-aneurysmal patients. The relatively low errors demonstrate that our approach can reliably identify aneurysmal patients and has potential utility in guiding clinical decision-making and surgical planning.

Table 4 Maximum Diameter Comparison

|  | Aortic Root | | | Ascending Aorta | | | Aortic Arch | | | Descending Aorta | | |
|---|---|---|---|---|---|---|---|---|---|---|---|---|
|  | GT | Pred | Error | GT | Pred | Error | GT | Pred | Error | GT | Pred | Error |
| **Non-Aneurysm** | 35.287 ±4.939 | 34.349 ±3.441 | 2.170 ±1.461 | 33.403 ±4.485 | 33.568 ±4.808 | 1.641 ±1.178 | 31.779 ±3.877 | 32.749 ±4.121 | 1.9447 ±1.736 | 25.553 ±3.201 | 27.077 ±5.938 | 2.004 ±4.128 |
| **Aneurysm** | 39.342 ±4.360 | 36.964 ±3.555 | 2.677 ±1.879 | 41.119 ±2.949 | 40.401 ±2.790 | 1.772 ±1.145 | 38.799 ±3.147 | 38.105 ±2.717 | 2.071 ±1.490 | 28.697 ±2.617 | 30.514 ±4.232 | 2.679 ±3.173 |
| **Total** | 37.315 ±5.081 | 35.656 ±3.735 | 2.423 ±1.702 | 37.261 ±5.412 | 36.985 ±5.208 | 1.706 ±1.163 | 35.289 ±4.979 | 35.427 ±4.399 | 2.008 ±1.619 | 27.125 ±3.319 | 28.795 ±5.435 | 2.341 ±3.697 |

In addition to the maximum cross-sectional diameter, stress serves as another representative biomechanical indicator of aortic condition. Figures 3–5 illustrate the comparison of peak stress across different anatomical regions under a blood pressure of 16 kPa. Figure 3 presents the results for non-aneurysmal patients, where the predicted peak stress in the ascending aorta averaged approximately 210 kPa, closely matching the ground-truth mean. The mean predicted peak stresses in other anatomical regions also showed close agreement with the ground truth, except in the descending aorta, where deviations were primarily attributed to the staircase effect observed in the predicted meshes.

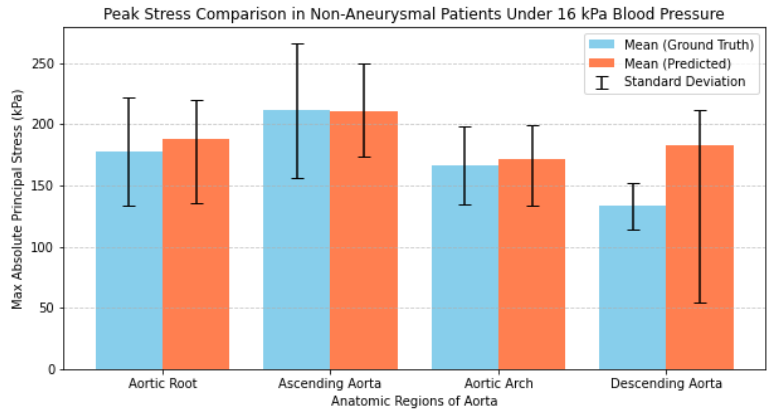

Figure 3 Peak Stress Comparison in Non-Aneurysmal Patients

Figure 4 presents the peak stress comparison results for aneurysmal patients. The predicted peak stress in the ascending aorta was approximately 265 kPa under a blood pressure of 16 kPa, closely matching the ground-truth mean. When comparing the ascending aorta region between non-aneurysmal and aneurysmal patients, the latter exhibited substantially higher peak stress values, consistent with the expected hemodynamic differences. This finding was accurately captured and validated by our proposed method.

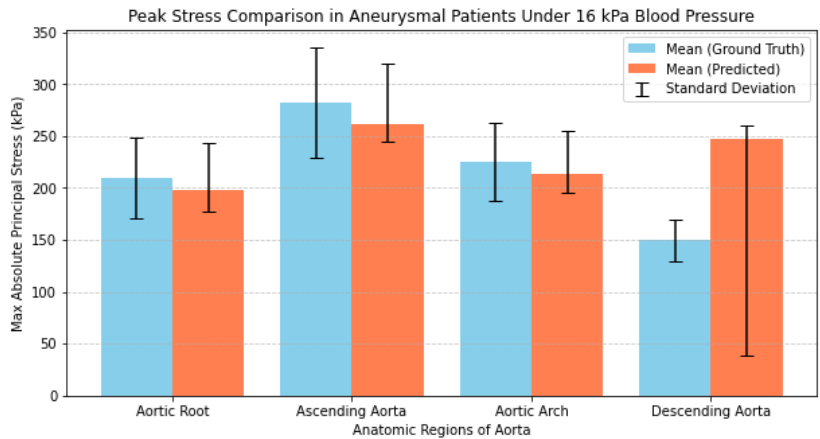

Figure 4 Peak Stress Comparison in Aneurysmal Patients

Figure 5 presents the peak stress comparison results for both non-aneurysmal and aneurysmal patients. The mean predicted peak stresses in the aortic root, ascending aorta, and aortic arch regions closely matched the ground-truth values.

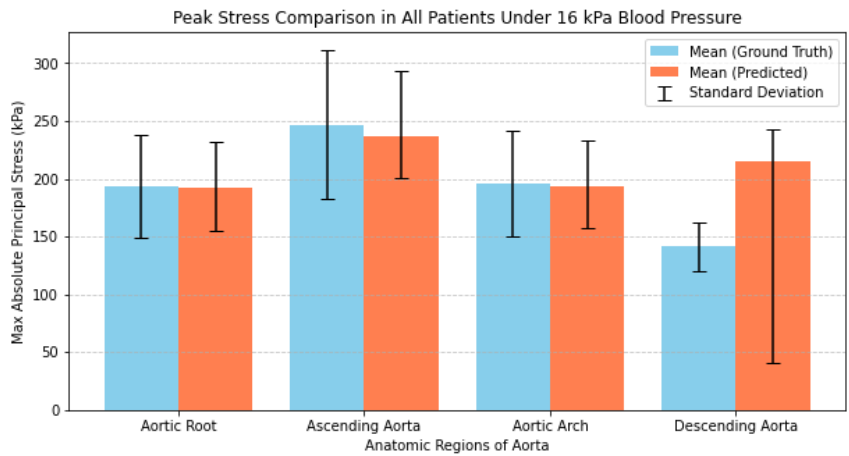

Figure 5 Peak Stress Comparison in All Patients

## 6. Conclusion

We proposed FEAorta, a fully automatic framework for patient-specific finite element analysis of the aorta directly from 3D CT images. By leveraging a deep learning–based image-to-mesh template deformation method, our method generates high-quality quadrilateral meshes with consistent mesh correspondence across different aortic regions. The framework demonstrates accurate geometric reconstruction, low simulation errors, and reliable stress estimation in both non-aneurysmal and aneurysmal patients. These capabilities enable the extraction of clinically relevant metrics, such as maximum cross-sectional diameters and regional stress distributions, supporting patient-specific assessment and surgical planning. In summary, FEAorta provides an efficient, accurate, and fully automated solution for comprehensive biomechanical analysis of the aorta.